\documentclass[aps,pra,twocolumn,nopacs,floatfix]{revtex4}
\usepackage{ulem}
\usepackage{epsfig}
\usepackage{graphicx}
\usepackage{dcolumn}
\usepackage{amsthm,amsmath}
\usepackage{comment}
\usepackage[colorlinks]{hyperref}
\usepackage{xcolor}
\usepackage{mathrsfs}
\usepackage{amsmath}
\usepackage{subcaption}
\usepackage[labelformat=simple]{subcaption}

\usepackage{morefloats}

\begin{document}


\title{Accurate estimate of $C_5$ dispersion coefficients of the alkali atoms interacting with different material media}

\author{$^1$Harpreet Kaur}
\author{$^1$Vipul Badhan}
\author{$^{1,2}$Bindiya Arora}
\email{bindiya.phy@gndu.ac.in}
\author{$^3$B. K. Sahoo}

\affiliation{$^1$Department of Physics, Guru Nanak Dev University, Amritsar, Punjab 143005, India}
\affiliation{$^2$Perimeter Institute for Theoretical Physics, Waterloo, Canada}
\affiliation{$^3$Atomic, Molecular and Optical Physics Division, Physical Research Laboratory, Navrangpura, Ahmedabad-380009, India}

\begin{abstract}
By inferring the dynamic permittivity of different material media from the observations and calculating dynamic electric dipole polarizabilties of the Li through Cs alkali atoms, precise values of $C_3$ coefficients were estimated in  Phys. Rev. A {\bf 89}, 022511 (2014) and Phys. Lett. A {\bf 380}, 3366 (2016). Since significant contribution towards the long range van der Waals potential is given by the quadrupole polarization effects, we have estimated the $C_5$ coefficients in this work arising from the quadrupole polarization effects of all the alkali atoms interacting with metal (Au), semiconductor (Si) and four dielectric materials (SiO$_2$, SiN$_x$, YAG and sapphire). The required dynamic electric quadrupole (E2) polarizabilities are evaluated by calculating E2 matrix elements of a large number of transitions in the alkali atoms by employing a relativistic coupled-cluster method. Our finding shows that contributions from the $C_5$ coefficients to the atom-wall interaction potentials are pronounced at short distances (1$-$10 nm). The $C_3$ coefficients of Fr atom interacting with the above material media are also reported. These results can be useful in understanding the interactions of alkali atoms trapped in different material bodies during the high-precision measurements. 
\end{abstract}

\maketitle

\section{Introduction}
Dispersion coefficients due to van der Waals (vdW) interactions between atoms and material walls have gained significant interest in the last two decades \cite{PhysRevA.70.053619} after their numerous applications in physisorption \cite{PhysRevB.8.5484,PhysRevA.104.012806}, storage \cite{PhysRevB.71.235401}, nano electromechanical systems \cite{bhojawala2017effect}, quantum reflection \cite{PhysRevA.78.042901}, atomic clocks \cite{PhysRevLett.103.133201}, atomic chips \cite{PhysRevLett.84.4749}, atom vapour sensors \cite{PhysRevA.100.022503} and so on.
The attractive potential between an atom and wall arises from quantum fluctuations at the zero contact point due to resonant coupling of virtual photons emitted from the atom with different electromagnetic modes of the surface of the wall \cite{bordag2009advances}. 
This phenomenun can be described by non-pairwise additive Lifshitz theory \cite{LIFSHITZ1992329}.
However, often crude approximations  have been  made in this theory for simplicity by considering only the dipole polarization effects due to their predominant contributions. Following the perturbation theory analysis, the atom-wall interaction potentials can be expressed as a sum of contributions from multipole-polarizability (i.e. dipole, quadrupole, octupole, etc.) effects of atoms \cite{Dalgarno_1967}. 

It has been pointed out that the corrections to the total potential due to multipole polarizations in atom-wall systems must be taken into account in the vicinity of physisorption rendered by the vdW interactions \cite{PhysRevB.35.9030}. 
Leibsch investigated the importance of the quadrupole contributions of atomic properties in the determination of atom-metal attractive interaction potentials and found 5-10\% enhancements in the contributions to the interaction potentials using the density functional theory (DFT) \cite{PhysRevB.35.9030}. 
For atoms placed closed to surfaces, some of the selective quadrupole resonances could play pivotal roles to enhance the atom-surface interaction significantly such that their contributions to the potentials can be higher than the dipole component contributions as noted by Klomiv \textit{et al.} using an analytical analysis~\cite{PhysRevA.62.043818}.
The dispersion coefficients arising from dipole ($C_3$) and quadrupole ($C_5$) interactions with the material walls were inspected by Tao \textit{et al.} between different atoms and metal surfaces using the DFT method and showed that $C_5$ term makes about 20\% contribution to the long range part~\cite{PhysRevLett.112.106101}.
 There are many other works that highlight the importance of higher-order multipole contributions to the atom-wall interaction potentials \cite{PhysRevB.81.233102,doi:10.1063/1.4940397,doi:10.1063/1.2190220,JIANG1984281,HUTSON1986L775,PhysRevA.81.052507}, whereas Lach \textit{et al.}~\cite{PhysRevA.81.052507} have provided a more accurate description of the vdW potentials for the interactions of atoms with the surfaces of perfect conductors and dielectrics materials by taking into account contributions from the dipole, quadrupole,  octupole and hexapole polarizabilities of the atoms within the framework of Lifshitz theory.

In the past decade, alkali atoms have been used to understand the behaviour of the vdW interactions with different materials using various theoretical and experimental techniques due to their fairly simple electronic configuration~\cite{PhysRevA.100.022503,sargsyan2020investigation,laliotis2021atom,SUN2015668,PhysRevA.92.052706}.
To gain the insight into the importance of quadrupole polarizability contributions from the alkali atoms towards their interaction potential with different material walls,  we have evaluated the vdW dispersion coefficients arising due to the dipole term ($C_3$) and next higher-order quadrupole term ($C_5$) for all alkali-metal atoms with different materials including metal, semiconductor and dielectrics over an arbitrary range of separation distance. 
Our work is in accordance with the previous studies that indicated the dominance of quadrupole polarization effects evaluated by other methods \cite{PhysRevB.35.9030,PhysRevA.62.043818,PhysRevLett.112.106101}. Particularly, we have probed the range of separation distance for which quadrupole effects are more significant. The $C_3$ and $C_5$ coefficients depend upon the polarizabilities of atoms and permittivity of the material walls at imaginary frequencies.
The accuracy of these coefficients can be achieved by using the appropriate methods to calculate these properties. We have used a relativistic all-order (AO) method to calculate the polarizabilities of the alkali atoms and Kramers-Kronig relation is used to determine the permittivity of materials at imaginary frequencies. Using these vdW coefficients, we have computed and investigated the potential curves for considered atom-wall systems.

In the following sections, we have provided brief theory related to the interaction potential at arbitrary separation, the method of evaluation of required properties of materials and atom in Sec.~\ref{sec3}, results and discussion in Sec.~\ref{sec4} and finally concluded our work in Sec.~\ref{sec5}. Unless stated otherwise, atomic units (a.u.) are being used through out the manuscript. 

\section{Theory}\label{sec2}

The exact theory for the calculation of vdW interaction potential between an atom and material surface has been given in Ref.~\cite{PhysRevA.81.052507}. Here, we give only a brief outline of the expressions for the atom-wall vdW interaction potentials due to multipole dispersion coefficients. The general expression of total attractive interaction potential ($U_{total}$) arising from the fluctuating multipole moments of an atom interacting with its image in the surface is given by~\cite{HUTSON1986L775}
\begin{equation}
U_{total}(z) = U_{d}(z) + U_{q}(z) + ...,
\end{equation}
where $U_{d}$, $U_{q}$, and so on are the contributions from the dipole, quadrupole etc. contributions and $z$ is the separation distance between atom and wall in nm. Due to predominant nature of dipole component, often $U_{total}(z)$ is approximated as $U_{d}(z)$, but we also estimate contributions from $U_{q}(z)$ in this work. In terms of permittivity values of the material and dynamic polarizabilities of the atoms, we can express \cite{osti_4359646,lach2010noble,PhysRevA.81.052507} 
\begin{eqnarray}\label{Ud}
& & U_{d}(z) = -\frac{\alpha_{fs}^3}{2\pi}\int_0^{\infty} d\omega \omega^3 \alpha_d(\iota\omega)
\nonumber \\
& & \times \int_1^{\infty} d\chi e^{2\chi\alpha_{fs}\omega z} H(\chi,\epsilon_r(\iota\omega))
\end{eqnarray}
and 
\begin{eqnarray}\label{Uq}
& & U_{q}(z) = -\frac{\alpha_{fs}^5}{12\pi}\int_0^{\infty} d\omega \omega^5 \alpha_q(\iota\omega)
\nonumber \\
& & \times \int_1^{\infty} d\chi e^{2\chi\alpha_{fs}\omega z}(2\chi^2 -1) H(\chi,\epsilon_r(\iota\omega)) .
\end{eqnarray}
In the above two expressions, $\alpha_{fs}$ is the fine-structure constant, $\chi$ is the Matsubara frequency, $\alpha_{d}(\iota\omega)$ and $\alpha_{q}(\iota\omega)$ are the dynamic dipole and quadrupole polarizabilities of the ground state of the considered atom at imaginary frequencies. The expression of function $H(\chi,\epsilon(\iota\omega))$ is given by~\cite{PhysRevA.89.022511}
\begin{equation}
H(\chi,\epsilon) = (1-2\chi^2)\frac{\chi'-\epsilon_r\chi}{\chi'+\epsilon_r\chi} + \frac{\chi'-\chi}{\chi'+\chi},
\end{equation} 
where $\chi' = \sqrt{\chi^2+\epsilon_r-1}$ and $\epsilon_r$ is the real part of dynamic permittivity of the material wall at imaginary frequency . Approximating the total potential till quadrupole effects, at short distances ($z \rightarrow 0$), the preceding formulas are now given by
\begin{equation}
U_{total}(z) = -\frac{C_3}{z^3} -\frac{C_5}{z^5}, 
\end{equation} 
where $C_3$ and $C_5$ coefficients are defined as
\begin{equation}\label{eqc3}
C_3 = \frac{1}{4\pi}\int_0^\infty d\omega \alpha_d (\iota\omega)\frac{\epsilon_r(\iota\omega)-1}{\epsilon_r(\iota\omega)+1},
\end{equation}
and
\begin{equation}\label{eqc5}
C_5 = \frac{1}{4\pi}\int_0^\infty d\omega \alpha_q (\iota\omega)\frac{\epsilon_r(\iota\omega)-1}{\epsilon_r(\iota\omega)+1}.
\end{equation}

\section{Method of evaluation}\label{sec3}

As mentioned in the previous section, evaluation of the $C_3$ and $C_5$ coefficients require knowledge of $\epsilon_r(\iota \omega)$ and $\alpha(\iota \omega)$ of the material media and atoms, respectively. The real part of permittivity at imaginary frequency $\epsilon_r(\iota \omega)$ values cannot be obtained experimentally, but their values can be inferred from the imaginary part of permittivity at real frequencies using the Kramers-Kronig relations. Similarly, accurate determination of dynamic values of $\alpha_d$ and $\alpha_q$ values at the imaginary frequencies are challenging in the {\it ab initio} approach. However for alkali atoms, these can be evaluated very accurately using the sum-over-states approach. Below, we discuss evaluation procedures of $\epsilon_r(\iota \omega)$ and $\alpha(\iota \omega)$.

\subsection{Dynamic electric permittivity}\label{3a}

The imaginary part of dynamic electric permittivity $\epsilon_i(\omega)$ can be given by
\begin{equation}
\epsilon_i(\omega) = 2 n(\omega)\kappa(\omega) ,
\end{equation}
where $n(\omega)$ and $\kappa(\omega)$ are the refractive indices and extinction coefficients of the materials at the real frequencies, respectively. Discrete $n(\omega)$ and $\kappa(\omega)$ values of the considered material media for a wide range of frequencies are tabulated in the Handbook on optical constants of solids by Palik~\cite{palik1998handbook}. Using these values, we have extrapolated values of $\epsilon_i(\omega)$ for continuous frequencies for a large range. Now using the Kramers-Kronig relation, we can express the real part of dynamic permittivities ($\epsilon_r(\iota\omega)$) at imaginary frequencies such that 
\begin{equation}
\epsilon_r(\iota\omega) = 1 + \frac{2}{\pi}\int_0^{\infty}d\omega' \frac{\omega' \epsilon_i(\omega')}{\omega^2 + \omega'^2}.
\end{equation}
For the case of semiconductor - Si and dielectrics - SiO$_2$, YAG, ordinary sapphire (oSap) and extraordinary sapphire (eSap), we have used the optical constants ranging from 0.1 eV to 10000 eV from Handbook of optical constants by Palik~\cite{palik1998handbook}. 
For the case of Au, the $\epsilon$ values at very small energies are very significant, hence in addition to the experimental values from Ref.~\cite{palik1998handbook}, we have extrapolated the values of real permittivity at imaginary frequencies using the Drude model for metals as
\begin{equation}
\epsilon_r(\iota\omega) = 1- \frac{\omega_p^2}{\omega(\omega + \iota\gamma)},
\end{equation}
where $\omega_p$ is the plasma frequency and $\gamma$ is the relaxation frequency. We have used $\omega_p$=9.0 eV and $\gamma$=0.035 eV as referred in ~\cite{lach2010noble,PhysRevA.61.062107}.
For the case of SiN$_x$, an amorphous dielectric material, we use Tauc–Lorentz model for amorphous materials~\cite{PhysRevA.71.053612} for estimating the electric permittivity at imaginary frequencies, the expression of which is given as follows
\begin{equation}
\epsilon_r(\iota\omega) = \frac{\omega^2 + (1+g_0)\omega_0^2}{\omega^2 + (1-g_0)\omega_0^2}, 
\end{equation} 
where the parameters $g_0 = 0.588$ and $\omega_0 = 0.005$ are the SiN$_x$'s response functions~\cite{PhysRevA.71.053612}.

\subsection{Dynamic polarizabilities}\label{3b}

We have already reported $\alpha_d(\iota \omega)$ values of the Li to Cs alkali atoms in our previous works \cite{PhysRevA.89.022511,KAUR20163366}. Here, we give $\alpha_d(\iota \omega)$ values of the Fr atom and the $\alpha_q (\iota \omega)$ values of all the alkali-metal atoms by evaluating them as given in the following procedures. 

Total electron correlation contributions to $\alpha_d(\iota \omega)$ and $\alpha_q(\iota \omega)$ of atomic states of the alkali atoms can be expressed as \cite{PhysRevA.91.012705}
\begin{equation}\label{alpha}
\alpha_l(\iota\omega)= \alpha_{l,core}(\iota\omega)+ \alpha_{l,vc}(\iota\omega) + \alpha_{l,val}(\iota\omega),
\end{equation} 
where $l = d$ corresponds to dipole polarizability and $l = q$ corresponds to quadrupole polarizability. Subscripts $core$, $vc$ and $val$ corresponds to core, valence-core and valence contributions, respectively, to the total polarizability. In the alkali atoms, $\alpha_{l,val}(\iota\omega)$ contributes predominantly followed by $\alpha_{l,core}(\iota\omega)$ and contributions from $\alpha_{l,vc}(\iota\omega)$ are negligibly small. These contributions are estimated in the following way.

To begin with, the electronic configuration of alkali atoms is divided into a closed-core and a valence orbital in order to obtain the mean-field Dirac-Fock (DF) wave function of the respective closed-shell ($|0_{c}\rangle$) using DF method. The mean-field wave functions of the atomic states of the alkali atoms are then defined by appending the respective valence orbital $v$ as 
\begin{equation}
|\phi_v\rangle = a_v^{\dagger}|0_{c}\rangle .
\end{equation}
Using these mean-field DF wave functions, we calculated the $vc$ contributions to the dipole and quadrupole polarizability using the following formula 
\begin{eqnarray}\label{pol}
& & \alpha_{l,vc}(\iota\omega)=\frac{2}{(2L+1)(2J_v+1)}
\nonumber \\
& & \times \sum_m^{N_c} \frac{(\mathcal{E}_m-\mathcal{E}_v)|\langle\psi_v||\textbf{O$_L$}||\psi_m\rangle_{DF}|^2}{(\mathcal{E}_m-\mathcal{E}_v)^2+\omega^2}.
	 \end{eqnarray}
where $J_v$ corresponds to total angular momentum  of the state. Similarly, the core contributions can be given by
\begin{eqnarray}\label{pol}
\alpha_{l,core}(\iota\omega)&=& \frac{2}{(2L+1)}
\nonumber \\
& & \times \sum_a^{N_c} \sum_{m}^B \frac{(\mathcal{E}_m-\mathcal{E}_a)|\langle\psi_a||\textbf{O$_L$}||\psi_m\rangle|^2}{(\mathcal{E}_m-\mathcal{E}_a)^2+\omega^2}, \ \ \ \ \
	 \end{eqnarray}
where the first sum for core orbitals is restricted from $a$ to total core orbitals $N_c$, second sum is restricted by involving intermediate states $m$ up to allowed bound states $B$ using the respective dipole and quadrupole selection rules, $L=1$ and $\textbf{O}_1 = \textbf{D}$ is for the dipole operator to give $\alpha_d$, $L = 2$ and $\textbf{O}_2 = \textbf{Q}$ is for quadrupole operator to give $\alpha_q$ and $\mathcal{E}_i$ is the DF energy of the state. We have adopted the random phase approximation (RPA) to evaluate the above expression to account for the core correlations \cite{PhysRevA.90.022511}.

\begin{table}[t]
\caption{\label{tab1} Contributions to the ground state dipole polarizability (in a.u.) of the Fr atom. Various contributions along with the absolute values of reduced E1 matrix elements contributing to the main part of the valence correlations are quoted explicitly. Tail, core and valence-core contributions are also given. Core contribution is estimated by the RPA method. Our final value is compared with the previously reported high-precision calculations.}
	\begin{center}
\begin{tabular}{p{2.5cm}p{1cm}p{1.5cm}}
\hline
\hline
& &  \\
Contribution  & E1  & $\alpha_{d}(0)$ \\ [0.5ex]
\hline 
& & \\
 Main   &    \\
$7S_{1/2}$	-	$7P_{1/2}$	&	4.277$^a$	&	109.36 \\
$7S_{1/2}$	-	$8P_{1/2}$	&	0.328	&	0.34 \\
$7S_{1/2}$	-	$9P_{1/2}$	&	0.11	&	0.03\\
$7S_{1/2}$	-	$10P_{1/2}$	&	0.056	&	0.01\\
$7S_{1/2}$	-	$11P_{1/2}$	&	0.034	&	0.003\\
$7S_{1/2}$	-	$12P_{1/2}$	&	0.024	&	0.001\\
$7S_{1/2}$	-	$7P_{3/2}$	&	5.898$^a$	&	182.77\\
$7S_{1/2}$	-	$8P_{3/2}$	&	0.932	&	2.68\\
$7S_{1/2}$	-	$9P_{3/2}$	&	0.435	&	0.51\\
$7S_{1/2}$	-	$10P_{3/2}$	&	0.270	&	0.18\\
$7S_{1/2}$	-	$11P_{3/2}$	&	0.191	&	0.09\\
$7S_{1/2}$	-	$12P_{3/2}$	&	0.145	&	0.05\\
 Tail &  & 1.101\\[0.5ex]
 Core  & & 20.4	\\[0.5ex]
$vc$ & & -1.0\\ [0.5ex]
  Total & & 316.61\\[0.5ex]
  Others &  & 317.8~\cite{PhysRevLett.82.3589}\\[0.5ex]
         &  & 313.7~\cite{PhysRevA.76.042504}\\[0.5ex] 
         &  & 325.8~\cite{PhysRevA.103.022802}\\[0.5ex]
 \hline
  \hline
	 \end{tabular}	
	 \end{center}
  $^a$Values are taken from Ref. \cite{PhysRevA.57.2448}.
	 \end{table}

The major contributions to the total dipole and quadrupole polarizabilities are provided by the $val$ contributions, hence it is important to calculate the same with accurate methods. We divide the $val$ contributions into two parts - main and tail. The main part corresponds to polarizability contributions by the low-lying dominant transitions responsible for very large polarizability contributions. For the evaluation of the main part of the $val$ contribution of total polarizability, we have employed the AO method to evaluate the accurate wave functions. These wave functions $|\psi_v\rangle$, with $v$ denoting the valence orbital, are represented using singles and doubles (SD) approximation of AO method as \cite{Safronova_1999} 
\begin{eqnarray}\label{eq12}
& &|\psi_v\rangle_{SD} = \left[1+ \sum_{ma}\rho_{ma}a_m^\dagger a_a +\frac{1}{2}\sum_{mrab}\rho_{mrab} a_m^\dagger a_r^\dagger a_b a_a\right.
\nonumber \\
& & \left.+ \sum_{m\neq v} \rho_{mv} a_m^\dagger a_v + \sum_{mla} \rho_{mrva}a_m^\dagger a_r^\dagger a_a a_v \right]|\phi_v\rangle ,
\end{eqnarray}				
where $a^\dagger$ and $a$ represent the second-quantization creation and annihilation operators, respectively, whereas excitation coefficients are denoted by $\rho$. The subscripts $m, r$ and $a, b$ refer to the virtual and core orbitals, respectively. $\rho_{ma}$ and $\rho_{mv}$ are the single whereas $\rho_{mrab}$ and $\rho_{mrva}$ are the double excitation coefficients. To take into account the important experimental contributions these \textit{ab initio} wave functions are modified by changing the valence excitation coefficient with modified $\rho_{mv}$ using the scaling procedure such that 
\begin{equation}
\rho'_{mv} = \rho_{mv}\frac{\delta E_v^{expt}}{\delta E_v^{theory}}.
\end{equation}
After obtaining wave functions of the considered states of alkali-metal atoms using AO method, we determine matrix elements with $k$  as intermediate state using the following expression~\cite{PhysRevA.40.2233}
\begin{eqnarray}\label{13}
	O_{L,vk} = \frac{\langle\psi_v|\textbf{O}_L|\psi_k\rangle}{\sqrt{\langle\psi_v|\psi_v\rangle \langle\psi_k|\psi_k\rangle}},
	\end{eqnarray}
where $O_{L,vk}$ corresponds to either dipole E1 or quadrupole E2 matrix elements depending on \textbf{D} or \textbf{Q} operators, respectively. Using these matrix elements, the final expression for the main part of the $val$ contribution to either the E1 or E2 polarizability at imaginary frequency is then given as
\begin{eqnarray}\label{pol}
& & \alpha_{l,Main}(\iota\omega)=\frac{2}{(2L+1)(2J_v+1)}
\nonumber \\
& & \times \sum_{m>N_c,m\neq v} \frac{(E_m-E_v)|\langle\psi_v||\textbf{O}_L||\psi_m\rangle|^2}{(E_m-E_v)^2+\omega^2}.
	 \end{eqnarray}
where the sum now restricted by entailing the intermediate states $m$ after $N_c$ and up to $I$. We have considered 10 - 12 E1 and E2 matrix elements for the dominant transitions of considered atoms using the AO method. For precise calculations, we use experimental energies $(E_i)$ from the National Institute of Science and Technology (NIST) database~\cite{NIST_ASD}. Contributions from the remaining high-lying states are referred as tail part and are evaluated as
\begin{eqnarray}\label{pol}
& & \alpha_{l,Tail}(\iota\omega)=\frac{2}{(2L+1)(2J_n+1)}
\nonumber \\
& & \times \sum_{m>I} \frac{(\mathcal{E}_m-\mathcal{E}_n)|\langle\psi_n||\textbf{O}_L||\psi_m\rangle_{DF}|^2}{(\mathcal{E}_m-\mathcal{E}_n)^2+\omega^2} ,
	 \end{eqnarray}
where $m>I$ means that states included in the main contribution evaluation are excluded here. Since the tail contributions are much smaller in comparison to the main part, we calculate them using the DF method.

\begin{table}
\caption{\label{c3fr}Calculated $C_3$ coefficients (in a.u.) of Fr with various material walls with explicit contributions from the core, $vc$, main and tail parts of its dipole polarizability.}
\begin{center}
\begin{tabular}{p{1.5cm}p{1cm}p{1cm}p{1cm}p{1cm}p{1cm}}
\hline \hline
& & & \\
& Core & $vc$ & Main & Tail & Total \\
\hline
& & & \\
Au & 0.899 & -0.033 & 1.928 & 0.024 & 2.818 \\[0.5ex]
Si & 0.644 & -0.025 & 1.643 & 0.020 & 2.282 \\[0.5ex]
SiO$_2$ & 0.394 & -0.014 & 0.776 & 0.010 & 1.166 \\[0.5ex]
SiN$_x$ & 0.482 & -0.019 & 1.173 & 0.011 & 1.650 \\[0.5ex]
YAG & 0.613 & -0.023 & 1.130 & 0.014 & 1.735 \\[0.5ex]
oSap & 0.658 & -0.024 & 1.162 & 0.015 & 1.812 \\[0.5ex]
eSap & 0.698 & -0.029 & 1.167 & 0.015 & 1.851 \\[0.5ex]
\hline
\hline
\end{tabular}
\end{center}
\end{table}

\begin{table*}[t]
\caption{\label{table1} Contributions to the ground state quadrupole polarizabilities (in a.u.) of the Li, Na, K, Rb, Cs and Fr atoms. Various contributions
along with the E2 matrix elements contributing to the main part of the valence correlations are quoted explicitly. Tail, core and valence-core contributions are also given. RPA value of core has been considered for total static value of $\alpha_q$. Final results are compared with the previously available values.}
	\begin{center}
\begin{tabular}{p{2.5cm}p{1cm}p{1.5cm}p{2.5cm}p{1cm}p{1.5cm}p{2.5cm}p{1cm}p{1.5cm}}
\hline
\hline
& & & & & & & & \\
\multicolumn{3}{c}{Li}  & \multicolumn{3}{c}{Na} &  \multicolumn{3}{c}{K} \\
\hline
& & & & & & & & \\
Contribution  & E2  & $\alpha_{q}(0)$ & Contribution  & E2  & $\alpha_{q}(0)$	& Contribution & E2 &  $\alpha_{q}(0)$ \\ [0.5ex]

 Main   &   &  &	 Main &   &	&	 Main &   &  \\
 
$2S_{1/2}$	-	$3D_{3/2}$	&	17.340  &	421.9	&	$3S_{1/2}$	-	$3D_{3/2}$ 	&	19.79	&	589 &	$4S_{1/2}$	-	$3D_{3/2}$	&	30.68	&	1919 \\

$2S_{1/2}$	-	$4D_{3/2}$	&	7.281	&	64	&	$3S_{1/2}$	-	$4D_{3/2}$	&	7.783	&	77.0	&	$4S_{1/2}$	-	$4D_{3/2}$	&	4.04	&	26	\\

$2S_{1/2}$	-	$5D_{3/2}$	&	4.301	&	21	&	$3S_{1/2}$	-	$5D_{3/2}$	&	4.466	&	23.64	&	$4S_{1/2}$	-	$5D_{3/2}$	&	0.45	&	0.3 \\

$2S_{1/2}$	-	$6D_{3/2}$	&	2.956	&	9	&	$3S_{1/2}$	-	$6D_{3/2}$	&	3.021	&	10	&	$4S_{1/2}$	-	$6D_{3/2}$	&	0.32	&	0.1	\\

$2S_{1/2}$	-	$7D_{3/2}$	&	2.207	&	5	&	$3S_{1/2}$	-	$7D_{3/2}$	&	2.234	&	6	&	$4S_{1/2}$	-	$7D_{3/2}$	&	0.49	&	0.3\\	
	
$2S_{1/2}$	-	$8D_{3/2}$	&	1.74	&	3	&	$3S_{1/2}$	-	$8D_{3/2}$	&	1.746	&	4	&	$4S_{1/2}$	-	$8D_{3/2}$	&	0.51	&	0.3	\\

$2S_{1/2}$	-	$3D_{5/2}$	&	21.237	&	632.8	&	$3S_{1/2}$	-	$3D_{5/2}$	&	24.23	&	884	&	$4S_{1/2}$	-	$3D_{5/2}$	&	37.58	&	2879	\\

$2S_{1/2}$	-	$4D_{5/2}$	&	8.917	&	95.3	&	$3S_{1/2}$	-	$4D_{5/2}$	&	9.532	&	115.4	&	$4S_{1/2}$	-	$4D_{3/2}$	&	4.94	&	39\\
	
$2S_{1/2}$	-	$5D_{5/2}$	&	5.27	&	31	&	$3S_{1/2}$	-	$5D_{5/2}$	&	5.470	&	35.46	& 	$4S_{1/2}$	-	$5D_{5/2}$	&	0.55	&	0.4\\

$2S_{1/2}$	-	$6D_{5/2}$	&	3.620	&	14	&	$3S_{1/2}$	-	$6D_{5/2}$	&	3.700	&	16	&	$4S_{1/2}$	-	$6D_{5/2}$	&	0.39	&	0.2	\\

$2S_{1/2}$	-	$7D_{5/2}$	&	2.703	&	8	&	$3S_{1/2}$	-	$7D_{5/2}$	&	2.737	&	8	&	$4S_{1/2}$	-	$7D_{5/2}$	&	0.60	&	0.5 \\	
	
$2S_{1/2}$	-	$8D_{5/2}$	&	2.126	&	5	&	$3S_{1/2}$	-	$8D_{5/2}$	&	2.139	&	5	&	$4S_{1/2}$	-	$8D_{5/2}$	&	0.62	&	0.5 \\
 
  Tail & &  114 & Tail &  & 104 & Tail  & & 98\\[0.5ex]
 
 Core  & &  0.112 & Core & & 1.52 &	Core & & 16.27	\\[0.5ex]
 
 
vc & & 0 & vc & & 0 & vc & & 0\\[0.5ex]
 
  Total & &  1424  & Total & & 1879 & Total & & 4980\\[0.5ex]
  
  Others & & 1423~\cite{PhysRevA.103.022802} & Others & & 1895~\cite{PhysRevA.103.022802} &	Others & & 4962~\cite{PhysRevA.103.022802} \\[0.5ex]
  
   & & &  & & 1800~\cite{JIANG1984281} & & & \\[0.5ex]
   
   
 \hline
& & & & & & & & \\
\multicolumn{3}{c}{Rb}  & \multicolumn{3}{c}{Cs} &  \multicolumn{3}{c}{Fr} \\
\hline
& & & & & & & & \\
Contribution  & E2  & $\alpha_{q}(0)$ & Contribution  & E2  & $\alpha_{q}(0)$	& Contribution & E2 &  $\alpha_{q}(0)$ \\ [0.5ex]

 Main   &   &  & Main &   &	& Main &   &  \\	  
 
$5S_{1/2}$	-	$4D_{3/2}$	&	32.94	&		2461 &	$6S_{1/2}$	-	$5D_{3/2}$	&	33.33	&	3362	&	$7S_{1/2}$	-	$6D_{3/2}$	&	32.91	&	2930 \\


$5S_{1/2}$	-	$5D_{3/2}$	&	0.31	&	0.2	&	$6S_{1/2}$	-	$6D_{3/2}$	&	12.56	&	306	&	$7S_{1/2}$	-	$7D_{3/2}$	&	8.09	&	119	\\

$5S_{1/2}$	-	$6D_{3/2}$	&	2.23	&	8	&	$6S_{1/2}$	-	$7D_{3/2}$	&	7.91	&	106	&	$7S_{1/2}$	-	$8D_{3/2}$	&	5.79	&	53	\\	
	
$5S_{1/2}$	-	$7D_{3/2}$	&	2.08	&	6	&	$6S_{1/2}$	-	$8D_{3/2}$	&	5.442	&	46.7	&	$7S_{1/2}$	-	$9D_{3/2}$  &	4.169	&	26.4	\\

$5S_{1/2}$	-	$8D_{3/2}$	&	1.75	&	4	&	$6S_{1/2}$	-	$9D_{3/2}$  &	4.047	&	24.9	&	$7S_{1/2}$	-	$10D_{3/2}$	&	3.172	&	14.6	\\

$5S_{1/2}$	-	$9D_{3/2}$	&	1.47	&	3.0	&	$6S_{1/2}$	-	$10D_{3/2}$	&	3.173	&	15.0	&	$7S_{1/2}$	-	$11D_{3/2}$	&	2.521	&	9	\\	

$5S_{1/2}$	-	$4D_{5/2}$	&   40.37	&	3696	&	$6S_{1/2}$	-	$5D_{5/2}$	&	41.23	&	5111	&	$7S_{1/2}$	-	$6D_{5/2}$	&	40.98	&	4487	\\


$5S_{1/2}$	-	$5D_{5/2}$	&	0.33	&	0.2	&	$6S_{1/2}$	-	$6D_{5/2}$	&	14.76	&	423	&	$7S_{1/2}$	-	$7D_{5/2}$	&	8.73	&	138	\\

$5S_{1/2}$	-	$6D_{5/2}$	&	2.69	&	11	&	$6S_{1/2}$	-	$7D_{5/2}$	&	9.432	&	150	&	$7S_{1/2}$	-	$8D_{5/2}$	&	6.516	&	67	\\	
	
$5S_{1/2}$	-	$7D_{5/2}$	&	2.51	&	9	&	$6S_{1/2}$	-	$8D_{5/2}$	&	6.525	&	67.2	&	$7S_{1/2}$	-	$9D_{3/2}$   &	4.755	&	33.8	\\

$5S_{1/2}$	-	$8D_{5/2}$	&	2.12	&	6	&	$6S_{1/2}$	-	$9D_{3/2}$	&	4.865	&	36.1	&	$7S_{1/2}$	-	$10D_{5/2}$	&   3.641	&	19.2	\\

$5S_{1/2}$	-	$9D_{3/2}$	 &	1.78	&	5	&	$6S_{1/2}$	-	$10D_{5/2}$	&   3.819	&	21.7	&	$7S_{1/2}$	-	$11D_{5/2}$	&	2.905	&	12.0	\\
 
Tail & & 224	&  Tail & & 644 & Tail & & 478	\\[0.5ex]
 
Core  & &  35.35	& Core  & & 86.38 & Core  & & 125.18  \\[0.5ex]
 
$vc$ & & $\sim0$ & $vc$ & & $\sim0$ & $vc$ & & $\sim0$ \\[0.5ex]
 
Total & & 6469	&  Total &  & 10400 	& Total & & 8512\\[0.5ex]

Others & & 6485~\cite{PhysRevA.103.022802} & Others & & 10498~\cite{PhysRevA.103.022802} & Others & & 9225~\cite{PhysRevA.103.022802} \\[0.5ex]
  
& & & & & 10600~\cite{JIANG1984281}	& & &  \\[0.5ex]
   

 \hline
 \hline
   
	 \end{tabular}	   
	 \end{center}
	 \end{table*}

\section{Results}\label{sec4}

\subsection{$C_3$ coefficients of Fr atom}

Previously, we had calculated the $C_3$ coefficients for various material walls interacting with alkali atoms~\cite{PhysRevA.89.022511,KAUR20163366} except for Fr, so we are not repeating the results for other alkali atoms in the present work. Here we provide the $C_3$ coefficients only for Fr with a number of material walls. We give the static $\alpha_d(0)$ value along with reduced E1 matrix elements of Fr and compare with the other high-precision calculations in Table \ref{tab1}. 
We have taken the experimental values of E1 matrix elements of the dominant dipole transitions of Fr~\cite{PhysRevA.57.2448}. Other E1 matrix elements are calculated using the method given in Sec.~\ref{sec3}. 
Our value is in excellent agreement with value given by Derevianko \textit{et al.} who used  high precision experimental values for E1 matrix elements for the principal transitions and other E1 values by SD method~\cite{PhysRevLett.82.3589}. The value given by Safronova \textit{et al.} is evaluated using SD method which deviates from our value by around 1\%~\cite{PhysRevA.76.042504}.
Though the method opted by us is same as of Refs.~\cite{PhysRevLett.82.3589,PhysRevA.76.042504} to calculate the dipole polarizability of Fr  but we have also scaled the E1 matrix elements using experimental energies as explained in Sec.~\ref{sec3}. 
Recently, Smialkowski \textit{et al.} used a molecular MOLPRO package to evalaute the dipole polarizability of Fr which is overestimated and diverges by $\sim$3\% from our value~\cite{PhysRevA.103.022802}. Using reduced E1 matrix elements given in the same table, we have estimated the dynamic $\alpha_d(\iota\omega)$ values and used them to estimate different contributions to $C_3$ as given in Table~\ref{c3fr}. The dominant contributor of $C_3$ coefficients is the main part followed by core, tail and $vc$. The total value of $C_3$ coefficient differs from material to material.
Consequently, the various contributions have been added up to provide a final value of $C_3$ coefficient. The core is providing 27\%-38\% of the share of total value of $C_3$ which is in accordance with the work by Derevianko \textit{et al.}~\cite{PhysRevLett.82.3589} where they emphasized the sensitivity of core towards dipole $C_3$ values. On the other hand, the tail contribution is $\sim1\%$ of the total value.

\subsection{Quadrupole polarizabilities}
	  
To evaluate the $C_5$ coefficients, we require quadrupole polarizability of alkali atoms. In Table~\ref{table1}, we present the static values of quadrupole polarizability $\alpha_q(0)$ of the ground states of alkali-metal atoms and compared our resulted values with the available literature. We calculated the static polarizability by putting $\omega = 0$ in the Eq.~\ref{alpha}. Since E2 matrix elements are required for calculation of dynamic polarizability, therefore we have provided the matrix elements of the dominant E2 transitions in Table~\ref{table1} for all the alkali atoms. The breakdown of polarizability into the main, tail, core and $vc$ polarizabilities are also presented. The main part of valence polarizability provides the dominant contribution followed by tail and core polarizability. The $vc$ contributions for Li, Na and K are zero due to non-availability of $D$ orbitals in the core of these atoms whereas very insignificant contributions have been encountered for Rb, Cs and Fr. For the final value of total static polarizability, we have added the core polarizability values from RPA. We did not find experimental $\alpha_q(0)$ results for any alkali atom to compare our theoretical values with. However, in the same table, we have compared our results with the most recent work by Smialkowski \textit{et al.} where they calculated the static quadrupole polarizability of alkali atoms using MOLPRO package of \textit{ab initio} programs~\cite{PhysRevA.103.022802}. Our static value of quadrupole polarizability deviates from the values reported by Smialkowski \textit{et al.} by less than 1\% for Li to Cs alkali atoms  whereas for Fr, the discrepancy is about 8\%. In another work~\cite{JIANG1984281}, Jiang \textit{et al.} evaluated the dynamic quadrupole polarizability of Na and Cs using oscillator method. We believe that our values are much more reliable than the compared ones due to accurate calculations of the matrix elements evaluated using the AO method.  

\begin{figure}
\includegraphics[width=\columnwidth,keepaspectratio]{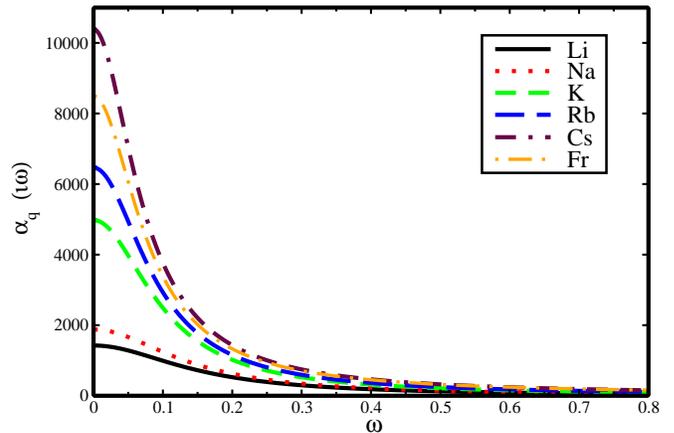}
\caption{Calculated dynamic quadrupole polarizability $\alpha_q$ (in a.u.) at imaginary frequencies of the alkali-atom metals.}
\label{qpol}
\end{figure} 
  
\begin{table*}
\caption{\label{fitting}Fitting parameters for the dynamic quadrupole polarizabilities of the alkali-metal atoms at imaginary frequencies.}
	\begin{center}
\begin{tabular}{p{2cm}p{1.5cm}p{1.5cm}p{1.5cm}p{1.5cm}p{1.5cm}p{1.5cm}}
\hline
\hline
\multicolumn{1}{c}{Parameter} & \multicolumn{6}{c}{Atom} \\
& Li & Na & K & Rb & Cs & Fr\\
\hline
$A$ & 1425.04 & 1879.5 & 4980.82 & 6470.66 & 10420.9 & 8514.71 \\
$B$ & 0.1296 & 0.1193& 0.2097 & 0.3640 & 1.1656 & 0.9549\\
$C$  & 42.0983 & 49.7813 & 97.0513 & 115.692 & 163.194 & 138.994 \\
\hline
\hline
\end{tabular} 
\end{center}
\end{table*}

\begin{table*}
\caption{\label{tablec3c5}Tabulated $C_5$ coefficients for the alkali-metal atoms with different material walls. Final results are compared with the previously available theoretical values.}
	\begin{center}
\begin{tabular}{p{2cm}p{1.5cm}p{1.5cm}p{1.5cm}p{1.5cm}p{1.5cm}p{1.5cm}p{1.5cm}p{1.5cm}}
\hline
\hline
\multicolumn{8}{c}{\textbf{Li}}  \\
\hline
 & Au & Si & SiO$_2$ & SiN$_x$ & YAG & oSap & eSap  \\[0.5ex]
Core  &
0.001 & 0.004 & 0.003 & 0.003 & 0.005 & 0.005 & 0.006 \\ [0.5ex]
vc & 
0 & 0 & 0 & 0 & 0 & 0 & 0 \\ [0.5ex]
Main &
17.657 & 14.870 & 7.292 & 10.808 & 10.827 & 11.181 & 11.320 \\ [0.5ex]
Tail & 
2.046 & 1.694 & 0.861 & 1.243 & 1.288 & 1.338 & 1.365 \\ [0.5ex]
Total  & 
19.710 & 8.156 & 16.568 & 12.054 & 12.120 & 12.524 & 12.691 \\ [0.5ex]
Ref. \cite{PhysRevLett.112.106101} &  19.15 & & & & & &\\
\hline
\\
\multicolumn{8}{c}{\textbf{Na}}  \\
\hline
 &
 Au & Si & SiO$_2$ & SiN$_x$ & YAG & oSap & eSap  \\[0.5ex]
Core & 
0.077 & 0.048 & 0.034 & 0.036  & 0.054 & 0.059 & 0.064 \\ [0.5ex]
vc &
0 & 0 & 0 & 0 & 0 & 0 & 0 \\ [0.5ex]
Main & 
22.610 & 19.081 & 9.308 & 13.846  & 13.801 & 14.242 & 14.401\\ [0.5ex]
Tail & 
1.762 & 1.464 & 0.738 & 1.072  & 1.103 & 1.144 & 1.165 \\ [0.5ex]
Total  & 
24.449 & 20.594 & 10.080 & 14.954 & 14.958 & 15.445 & 15.631 \\ [0.5ex]

Ref. \cite{JIANG1984281} &  25.2 & & & & & &\\
Ref. \cite{PhysRevLett.112.106101} &  22.48 & & & & & &\\
\hline
\\
\multicolumn{8}{c}{\textbf{K}}  \\
\hline
 & Au & Si & SiO$_2$ & SiN$_x$  & YAG & oSap & eSap  \\[0.5ex]
Core & 
0.702 & 0.465 & 0.308 & 0.347  & 0.486 & 0.530 & 0.571 \\ [0.5ex]
vc & 
0 & 0 & 0 & 0 & 0 & 0 & 0 \\ [0.5ex]
Main & 
47.036 & 39.962 & 19.117 & 28.784  & 28.151 & 28.972 & 29.146\\ [0.5ex]
Tail & 
1.873 & 1.539 & 0.792 & 1.132  & 1.188 & 1.236 & 1.264 \\ [0.5ex]
Total  & 
49.611 & 41.967 & 20.217 & 30.263  & 29.825 & 30.739 & 30.981 \\ [0.5ex]
Ref. \cite{PhysRevLett.112.106101} &  47.48 & & & & & &\\
\hline
\\
\multicolumn{8}{c}{\textbf{Rb}}  \\
\hline
 & Au & Si & SiO$_2$ & SiN$_x$  & YAG & oSap & eSap  \\[0.5ex]
Core & 
1.435 & 0.973 & 0.631 & 0.728  & 0.992 & 1.076 & 1.154 \\ [0.5ex]
vc & 
$\sim0$ & $\sim0$ & $\sim0$ & $\sim0$ & $\sim0$ & $\sim0$ & $\sim0$ \\ [0.5ex]
Main &
55.231 & 46.980 & 22.377 & 33.765  & 32.876 & 33.818 & 33.973\\ [0.5ex]
Tail & 
3.743 & 3.106 & 1.569 & 2.274 & 2.345 & 2.433 & 2.479 \\ [0.5ex]
Total  & 
60.410 & 51.059 & 24.577 & 36.766  & 36.213 & 37.327 & 37.606 \\ [0.5ex]
\hline
\\
\multicolumn{8}{c}{\textbf{Cs}}  \\
\hline
 &  Au & Si & SiO$_2$ & SiN$_x$  & YAG & oSap & eSap  \\[0.5ex]
Core &
3.277 & 2.287 & 1.442 & 1.712 & 2.252 & 2.431 & 2.591 \\ [0.5ex]
vc  &
$\sim0$ & $\sim0$ & $\sim0$ & $\sim0$ & $\sim0$ & $\sim0$ & $\sim0$ \\ [0.5ex]
Main & 
73.938 & 62.855 & 29.809 & 45.015  & 43.596 & 44.834 & 44.941\\ [0.5ex]
Tail & 
9.070 & 7.605 & 3.761 & 5.538  & 5.594 & 5.785 & 5.867 \\ [0.5ex]
Total  &
86.184 & 72.748 & 35.011 & 52.266  & 51.442 & 53.050 & 53.399 \\ [0.5ex]
Ref.~\cite{JIANG1984281} & 117 & & & & & & \\
\hline
\\
\multicolumn{8}{c}{\textbf{Fr}}  \\
\hline
 & Au & Si & SiO$_2$ & SiN$_x$  & YAG & oSap & eSap \\[0.5ex]
Core & 4.418 & 3.126 & 1.943 & 2.341  & 3.026 & 3.257 & 3.462 \\ [0.5ex]
vc & $\sim0$ & $\sim0$ & $\sim0$ & $\sim0$ &$\sim0$  & $\sim0$ & $\sim0$ \\ [0.5ex]
Main & 64.978 & 55.310 & 26.252 & 39.656  & 38.457 & 39.548 & 39.669\\ [0.5ex]
Tail & 7.169 & 5.987 & 2.985 & 4.369  & 4.448 & 4.606 & 4.680 \\ [0.5ex]
Total & 76.566 & 64.423 & 31.181 & 46.365  & 45.931 & 47.411 & 47.812 \\ [0.5ex]

\hline
\hline

\end{tabular} 
\end{center}
\end{table*}


\begin{figure*} 
\begin{subfigure}{0.45\textwidth}
\includegraphics[height=6cm,width=8.5cm]{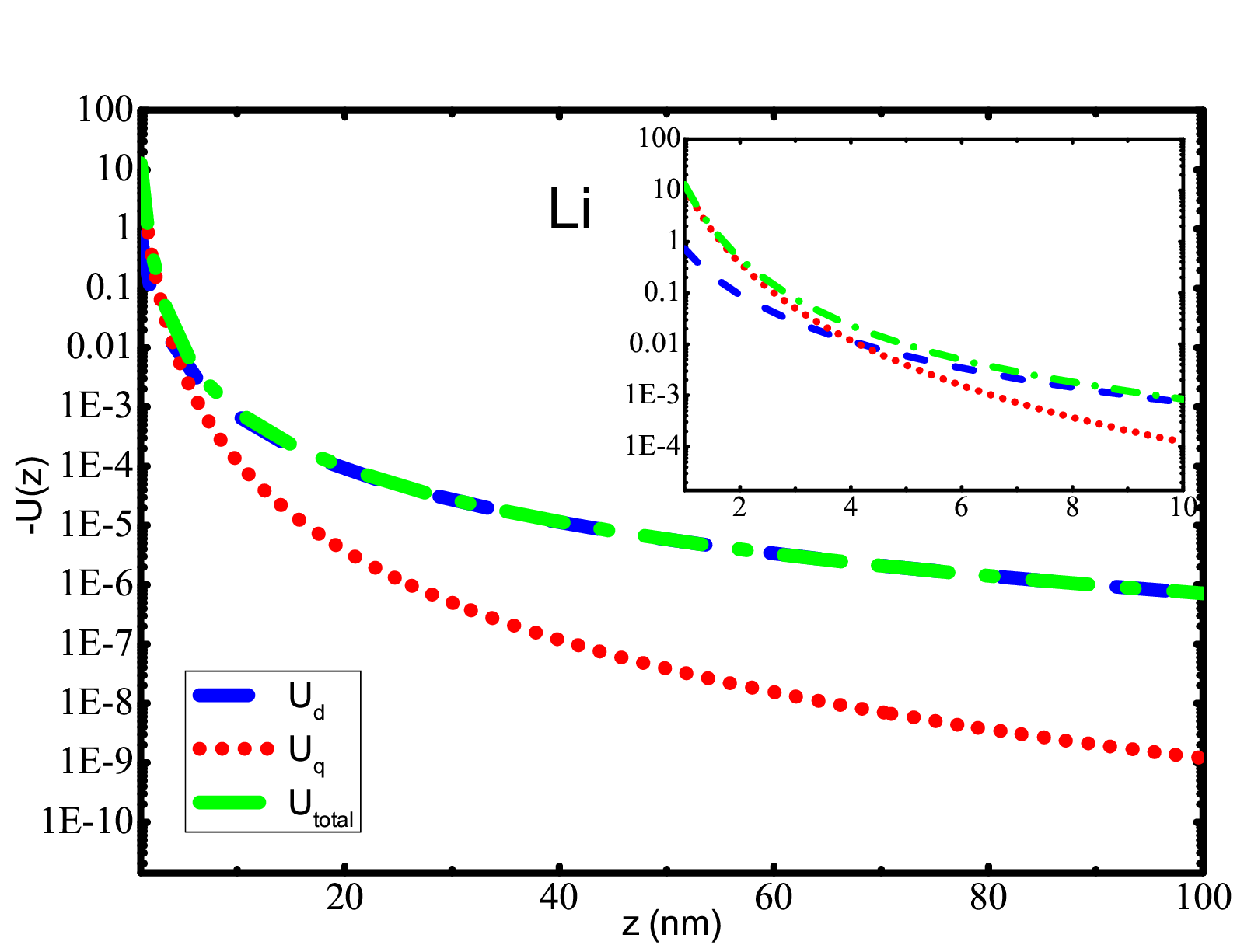}
\caption{Li}
\label{li}
\end{subfigure}
\begin{subfigure}{0.45\textwidth}
\includegraphics[height=6cm,width=8.5cm]{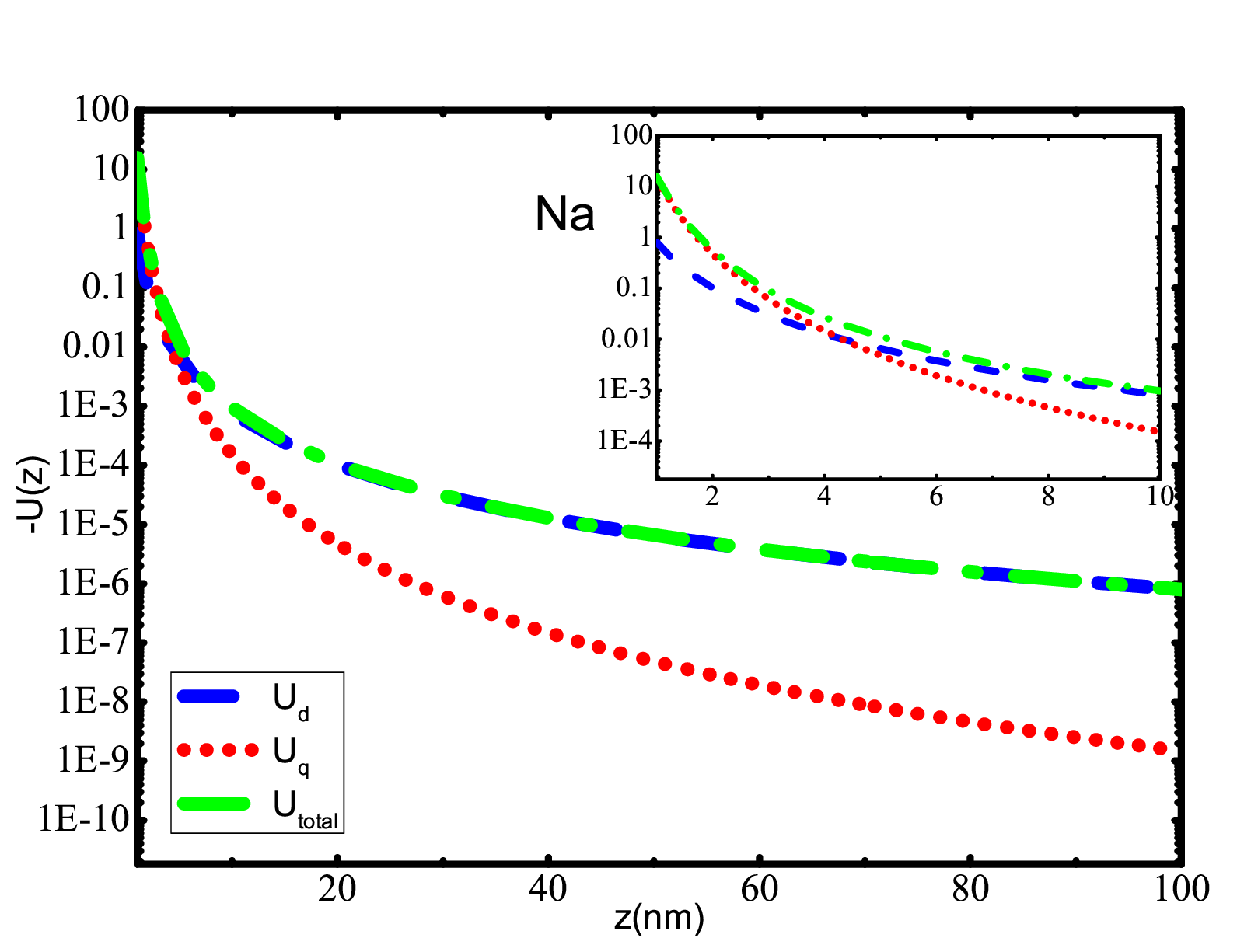}
\caption{Na}
\label{na}
\end{subfigure}
\begin{subfigure}{0.45\textwidth}
\includegraphics[height=6cm,width=8.5cm]{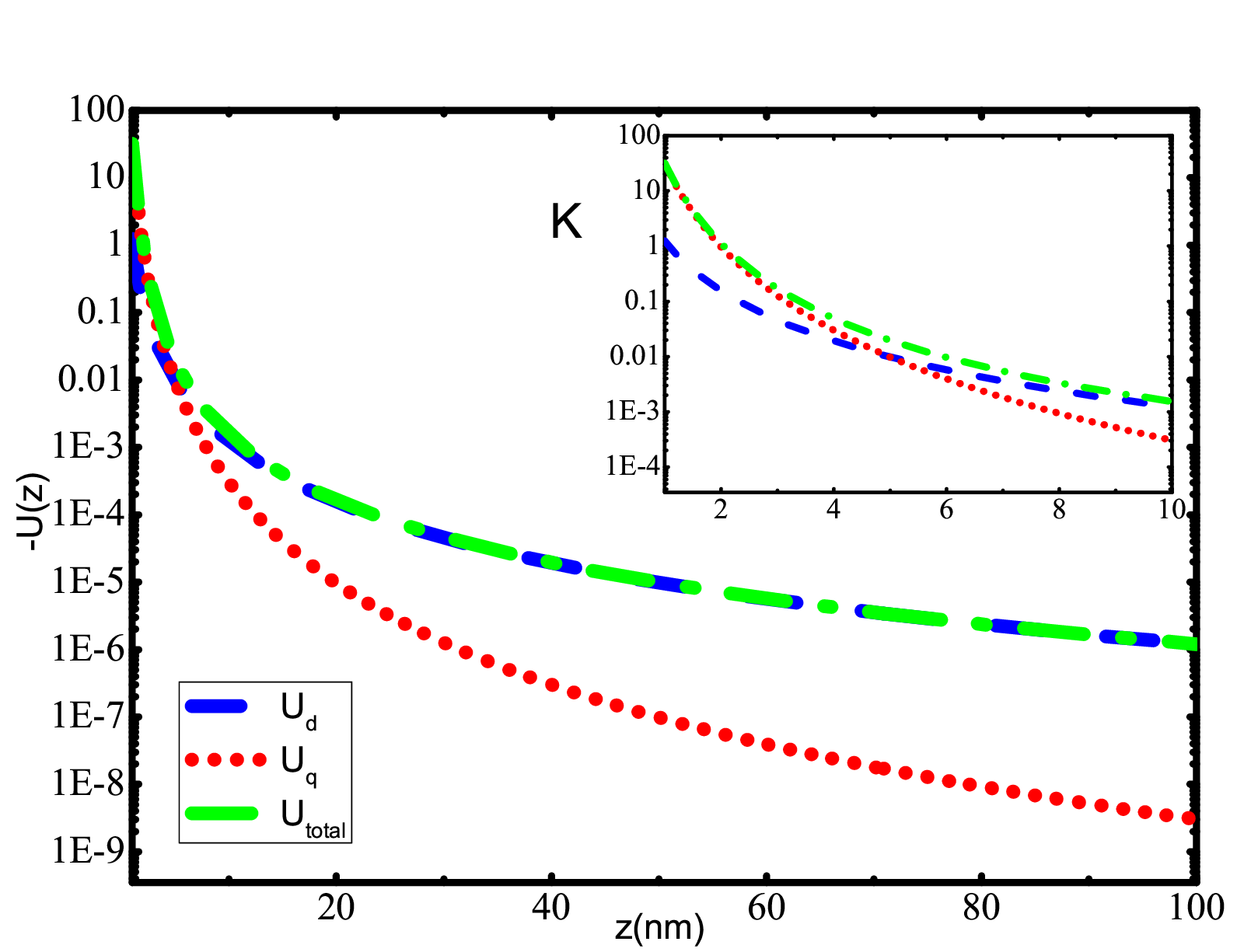}
\caption{K}
\label{k}
\end{subfigure}
\begin{subfigure}{0.45\textwidth}
\includegraphics[height=6cm,width=8.5cm]{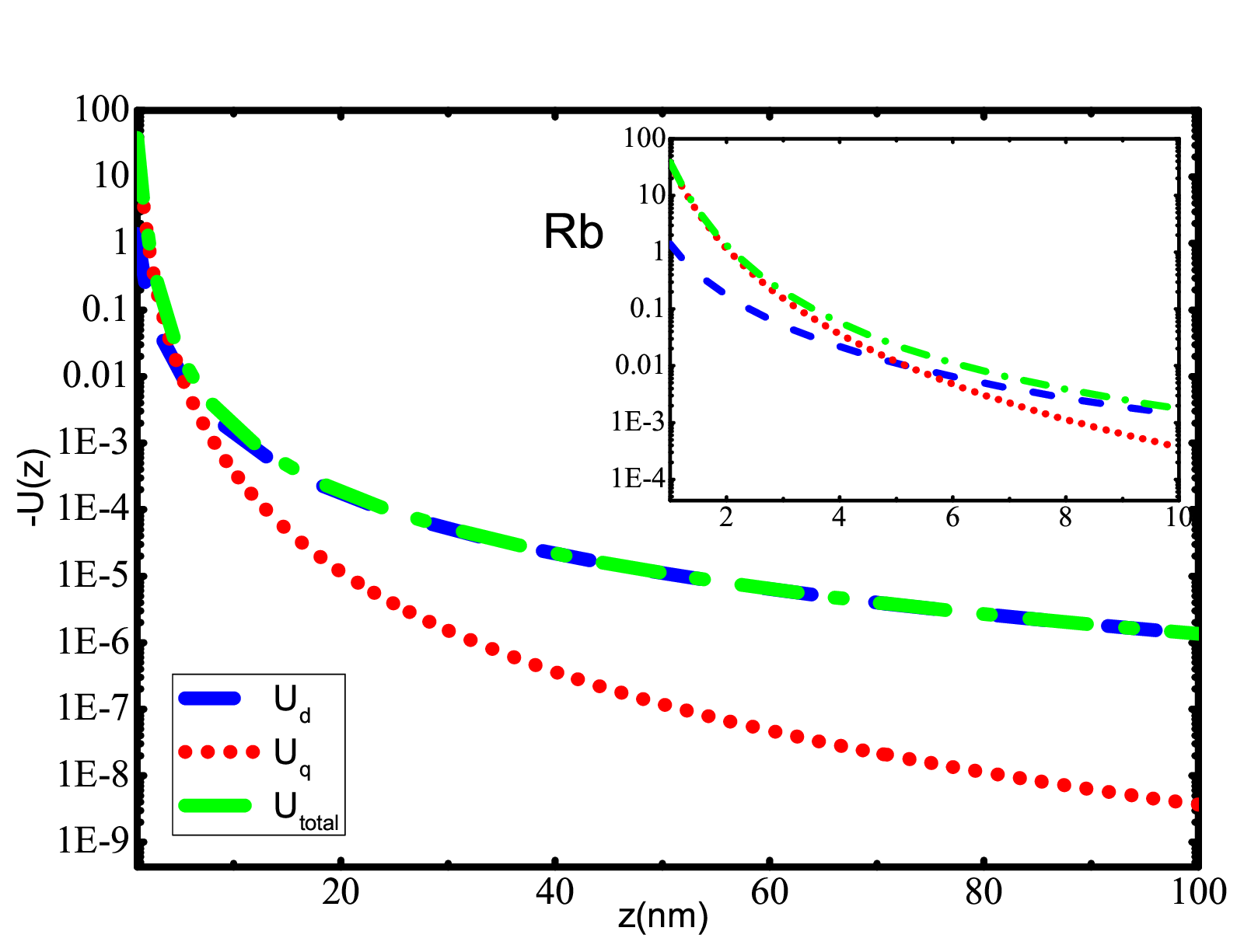}
\caption{Rb}
\label{rb}
\end{subfigure}
\begin{subfigure}{0.45\textwidth}
\includegraphics[height=6cm,width=8.5cm]{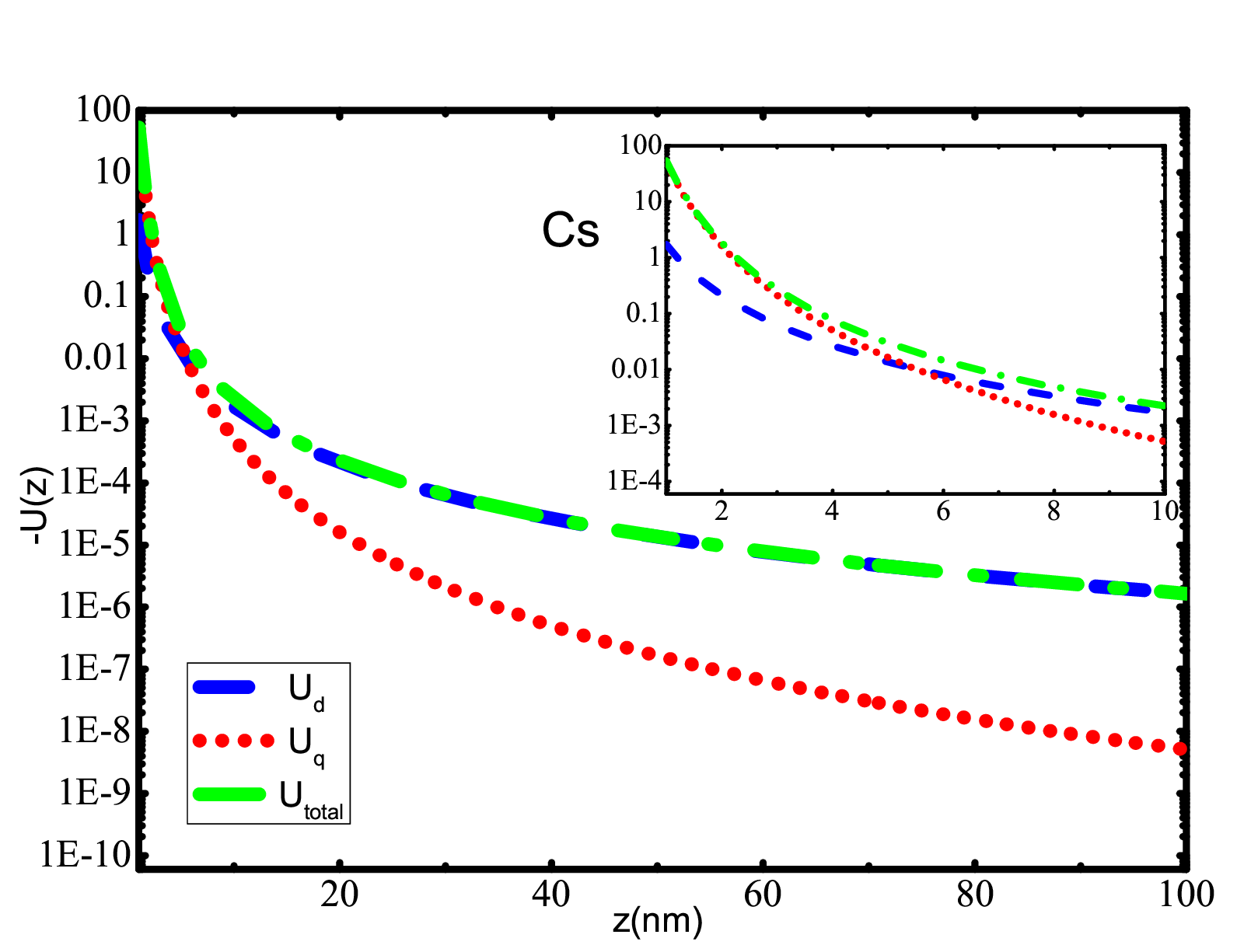}
\caption{Cs}
\label{cs}
\end{subfigure}\begin{subfigure}{0.45\textwidth}
\includegraphics[height=6cm,width=8.5cm]{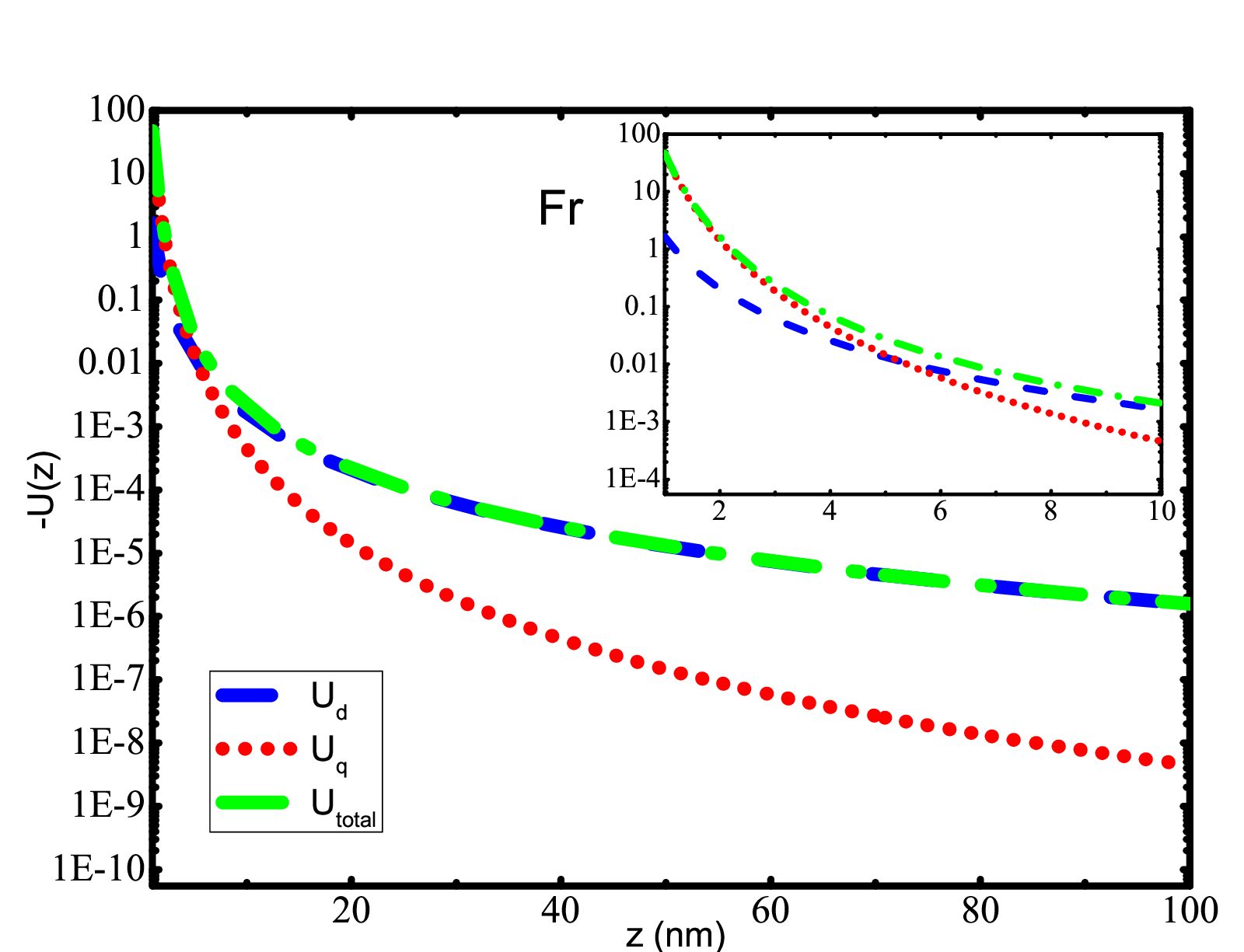}
\caption{Fr}
\label{fr}
\end{subfigure}
\caption{The vdW potential curves for interactions of the alkali-metal atoms with SiN$_x$ for $z=1-100$ nm. The insets of the graphs presents the same potential curves at $z=1-10$ nm.}
\label{potgraphs}
\end{figure*}
 
Using the E2 matrix elements, dynamic quadrupole polarizability  $\alpha_q(\iota\omega)$ of the alkali atoms over a range of frequencies has been calculated as presented in Fig.~\ref{qpol}. Since our static values are accurate, we believe that the dynamic values are also reliable. We find the static RPA and DF values for the core contributions are quite close, so we have estimated the dynamic values of core polarizabilities using the DF method without losing much accuracy. As the frequency increases the polarizability value decreases and reaches a small value beyond $\omega=1$ a.u.. This trend is seen for every atom considered in the present work.  Since these dynamic polarizability values can be important for experimental purposes, we have inferred these values at a particular frequency by providing a fitting model.  
In our previous work~\cite{dutt2020van}, we gave the fitting formula for dipole polarizability of alkali atoms at imaginary frequencies. Here, we have fitted the quadrupole polarizabilities of all the alkali atoms using the following fitting formula 
\begin{equation}
\alpha_q(\iota\omega) = \frac{A}{1 + B\omega + C^2\omega},
\end{equation}
where $A, B$ and $C$ are the fitting parameter given in Table ~\ref{fitting}.

\subsection{$C_5$ dispersion coefficients}

Table~\ref{tablec3c5} presents the calculated dispersion coefficients for all considered atoms due to $C_5$ contributions of polarizability in total potential interacting with different material walls. 
Using the resulted dynamic polarizability of considered atoms and dynamic permittivity values of material walls at imaginary frequencies, we have obtained the $C_5$ vdW dispersion coefficients by using Eq.~(\ref{eqc5}).  
We have used an exponential grid for solving the integration of the mentioned equation. In Table~\ref{tablec3c5}, core, $vc$, main and tail contributions of dispersion coefficients are given which are explicitly based on the corresponding contribution of polarizability. The final value of $C_5$ coefficient has been given by adding up all the contributions. The increasing size of the alkali atoms increases the values of individual contribution and total dispersion coefficients which is due to increasing polarizability of atom for any particular material wall.
Among the various contributions, the main part is the dominant contributor toward the total $C_5$ dispersion coefficient value, followed by tail, core and $vc$.
In Ref.~\cite{PhysRevLett.82.3589}, Derevianko \textit{et al.} emphasized that $C_3$ coefficients are sensitive towards the core contribution. However, for the case of $C_5$ coefficients, the core $C_5$ coefficients are much smaller and contribute atmost 5\% towards total $C_5$ value whereas the tail $C_5$ contributions are prominent.
The zero value of $vc$ contribution of $C_5$ coefficient for Li, Na and K is due to zero value of quadrupole polarizability.
After comparing the presented $C_5$ coefficients with $C_3$ coefficients that are already reported in our previous work ~\cite{PhysRevA.89.022511,KAUR20163366}, it can be observed that $C_3$ values are at least 25 times smaller than $C_5$ values for any particular system. The reason for this difference solely depends on the larger quadrupole polarizability of alkali atoms as compared to their dipole polarizability. Comparing the materials considered in the present work, the largest $C_5$ values have been observed for metal - Au, followed by semiconductor - Si and then dielectrics - sapphire, YAG, SiN$_x$ and SiO$_2$. 

We have also compared our values with the available theoretical vaues of $C_5$ coefficients. Jiang \textit{et al.} reported $C_5$ coefficients for Na and Cs with different materials including Au~\cite{JIANG1984281}. The reported value for Na-Au system is in close agreement with our value. But for the case of Cs-Au system, our value deviated from reported value by 35\%. The reason behind the discrepancies can be the method used for the calculation of dynamic polarizability and permittivity of atom and material, respectively. 
The oscillator method has been used to calculate the quadrupole polarizability of Na and Cs. This method overestimates the polarizability values, especially when the systems become heavier~\cite{Arora_2014}. It can also be seen from Table that the $\alpha_d(0)$ for Cs is not a reliable one. This could be one of the reasons behind the overestimated value of $C_5$ coefficient reported by Jiang \textit{et al.}.
The dynamic polarizability values of Na and Cs by Jiang \textit{et al.} have been evaluated by using oscillator method which gave a little deviated results from our values whereas the dynamic permittivity of Au has been evaluated using single frequency Lorentzian approximation, as a result of which the $C_5$ coefficients reported by them are exaggerated. 
In another report, Tao \textit{et al.} calculated the $C_5$ coefficients for Li, Na and K with Au using \textit{ab initio} DFT+vdW method~\cite{PhysRevLett.112.106101}. 
Though our values support the values reported in Ref.~\cite{PhysRevLett.112.106101}, the deviation of our values from the reported ones start increasing with increase in size of atom. As it is commonly known that the exchange correlation functional and nonlocal correlation energies are not treated properly in DFT method, we believe our values are accurate and more reliable than the values given by Tao \textit{et al.}.
	  
\subsection{Total vdW potentials}

The primary findings of the present work have been given in this section. Fig.~\ref{potgraphs} presents the potential curves due to dipole and quadrupole effects evaluated using Eqs. (\ref{Ud}) and (\ref{Uq}), respectively for alkali-metal atoms interacting with SiN$_x$ system.
Most of the experiments have been conducted with SiN$_x$ diffraction grating~\cite{perreault2005using,lepoutre2011atom,PhysRevA.80.062904,PhysRevLett.105.233202,PhysRevLett.83.1755}, so for demonstration purposes, we have chosen SiN$_x$ wall to observe the total potential curves with alkali atoms.
The total potential curve has been obtained till the first higher-order interaction of atom-wall system within the framework of Lifshitz theory. The individual dipole and quadrupole potential curve have also been  plotted in the same graphs. We have evaluated $U_d$ using our previous value of dipole polarizability~\cite{PhysRevA.89.022511,KAUR20163366}. It can be observed that quadrupole contribution gives very small contribution towards total potential. If we scrutinize these graphs at a very short separation distances, i.e., from $z = 1$ to $z = 10$ a.u., as presented in the insets of graphs~\ref{li}-~\ref{fr}, one can observe the overwhelming contribution provided by quadrupole contribution of the atom-wall potential. 
The quadrupole contribution is more dominant than dipole from 1 nm to 6 nm for all the alkali-metal atoms. 
As the separation distance increases, the long range dispersion interaction is completely imparted by dipole effect of polarizability of the atom as depicted in the figures. 
These results suggest that for a particular material the multipole effects can be quite significant if the separation distance is very small. Also, the multipole effect is much more effective and can be realized over larger separation when the atom or molecule considered is profoundly polarized. 
Similar curves can be obtained for the other materials that have been considered in this work. 

\section{Conclusion}\label{sec5}

We have investigated the quadrupole polarization effects of alkali atoms in the total atom-wall van der Waals interaction potentials. We probed the range of separation distance at which these quadrupole effects are dominant. For this, we considered both dipole and quadrupole induced interactions of atoms with various material walls within the framework of Lifshitz theory. The potential curves depict that quadrupole polarization effects of alkali atoms in total atom-wall potential are quite significant when the separation distance between atom and material wall is ranging from 1 - 10 nm. Beyond this range, the quadrupole contributions start declining, resulting in an attractive potential entirely due to the dipole polarization effects. 
Also, at significantly shorter distances, the attraction due to quadrupole polarization of the alkali atom increase with increase in the size of the atom suggesting quadrupole effects can be dominant when an atom has more tendency to get polarized. The obtained results could be useful in high precision experiments for studying van der Waals interactions at smaller distances very close to the surfaces.  

\section{Acknowledgement}
Research at Perimeter Institute is supported in part by the Government of Canada through the Department of Innovation, Science and Economic Development and by the Province of Ontario through the Ministry of Colleges and Universities.


\end{document}